\documentclass[twocolumn,showpacs,preprintnumbers,amsmath,amssymb,superscriptaddress]{revtex4-1}

\usepackage{graphicx}% Include figure files
\usepackage{dcolumn}% Align table columns on decimal point
\usepackage{bm}% bold math

\begin{document}

\title{Comment on ``Heat dissipation in atomic-scale junctions'' {\normalfont[Nature {\bf 498}, 209 (2013)]}}

\author{Y. Apertet}\email{yann.apertet@gmail.com}
\affiliation{Lyc\'ee Jacques Pr\'evert, F-27500 Pont-Audemer, France}
\affiliation{Institut d'Electronique Fondamentale, Universit\'e Paris-Sud, CNRS, UMR 8622, F-91405 Orsay, France}
\author{H. Ouerdane}
\affiliation{Universit\'e Paris Diderot, Sorbonne Paris Cit\'e, Institut des Energies de Demain (IED), 75205 Paris, France}
\affiliation{Laboratoire CRISMAT, UMR 6508 CNRS, ENSICAEN et Universit\'e de Caen Basse Normandie, 6 Boulevard Mar\'echal Juin, F-14050 Caen, France}
\author{C. Goupil}
\affiliation{Universit\'e Paris Diderot, Sorbonne Paris Cit\'e, Institut des Energies de Demain (IED), 75205 Paris, France}
\affiliation{Laboratoire CRISMAT, UMR 6508 CNRS, ENSICAEN et Universit\'e de Caen Basse Normandie, 6 Boulevard Mar\'echal Juin, F-14050 Caen, France}
\author{Ph. Lecoeur}
\affiliation{Institut d'Electronique Fondamentale, Universit\'e Paris-Sud, CNRS, UMR 8622, F-91405 Orsay, France}
\date{\today}

\begin{abstract}
We discuss the recent results of Lee et al. [Nature {\bf 498}, 209 (2013)] using the notion of \emph{thermoelectric convection}. In particular, we highlight the fact that this contribution to the thermal flux is not a dissipative process.
\end{abstract}


\maketitle

In a recent article \cite{Lee2013}, Lee and coworkers study the heat fluxes flowing inside a molecular junction when a difference of electrical potentials is applied. They focus in particular on the repartition of the heat between both side of the junction: they observe an asymmetry of the outcoming heat between each side for a single molecule junction. The authors thus conclude that the dissipation inside the junction is asymmetric in this case. They also suggest that this result is related to the mesoscopic nature of the system. We believe however that the interpretation of these remarkable experimental results should be reconsidered. In particular the term \emph{dissipation} seems inappropriate to qualify the total heat flux. Indeed, while Joule heating is of course dissipative, the asymmetric fraction of the heat flux should rather be related to convective transport, which is reversible and thus not associated to dissipation.

\section*{Ioffe's description of thermoelectric conversion}

In order to highlight the importance of the convective heat flux, we use for sake of simplicity the linear model proposed by Ioffe to describe the behavior of thermoelectric module \cite{Ioffe1957}. In this model the heat fluxes at each side of the molecular junction may be expressed, using the notations of Ref.~\cite{Lee2013}, as: 
\begin{eqnarray}\label{eq1}
\nonumber
Q_P &=& - S T_P I - K \left(T_P - T_S\right) + \frac{1}{2G} I^2 \\
&&~\\
\nonumber
Q_S &=& S T_S I + K \left(T_P - T_S\right) + \frac{1}{2G} I^2
\end{eqnarray}
\noindent where $S$, $K$ and $G$ are respectively the Seebeck coefficient, the thermal conductance and the electrical conductance of the junction, $T_P$ and $T_S$ are the temperature at the edges of the junction and $I$ is the electrical current flowing inside the system.

Both expressions are a summation of three different contributions: the first term is related to the heat transfer by the global movement of the charge carriers, and is thus denoted \emph{convective transfer} \cite{Thomson1856}, the second is the classical heat conduction described by Fourier's law and the last one is associated to the Joule heating inside the system. The convection process depends of course on the Seebeck coefficient $S$ since the heat transported by the charge carrier is $S T / e$, with $T$ the local temperature and $e$ the electrical charge of a carrier. While both heat conduction and Joule heating are independent of the direction of the electrical current, this is not the case for the thermoelectric convection.

In Ref.~\cite{Lee2013}, as the thermal resistance of the junction is much higher than that of the contacts (at least 100 times higher according to the supplementary materials of Ref.~\cite{Lee2013}), one may assume that $T_P \approx T_S = T$, $T$ being the ambient temperature. In this case the conductive part of the thermal flux is negligible and the contribution of the convective process to $Q_P$ is exactly the opposite of its contribution to $Q_S$.

It is possible to recover the expressions derived in Ref.~\cite{Lee2013} using Ioffe's model; To this purpose, it is necessary to express $I$ as a function of the applied voltage $V$:
\begin{equation}
I = G \left[S\left(T_P - T_S\right) - V\right]
\end{equation}
\noindent Since $T_P = T_S$, the electromotive force due to thermoelectric effects vanishes, so $I = - GV$. Eq.~(2) of Ref.~\cite{Lee2013} is recovered by replacing this expression in Eq.~(\ref{eq1}). It is thus obvious that the asymmetry between the two heat fluxes observed by Lee and coworkers stems from the presence of the convective transport of heat. Note that this result holds for any thermoelectric device independently of its size. In particular, it is not restricted to mesoscopic systems.
Considering the substrate as the thermal reservoir of interest, it is possible to identify the behavior of the molecular junction to the one of a classical thermoelectric module : When the applied voltage is positive, $Q_S$ is negative which corresponds to the refrigerator working condition while when the applied voltage is negative, $Q_S$ is positive which corresponds to the heat pump working condition. When the voltage becomes high in absolute value, the Joule heating overcomes the convective flux and the molecular junction behaves mainly as a dissipative system.

\section*{On the notion of dissipation}

In Ref.~\cite{Lee2013}, Lee and coworkers focus on the repartition of dissipated heat between both ends of the system. They make for this purpose the hypothesis that the heat fluxes are fully related to dissipation. However it is not the case: indeed the convective flux should not be associated to dissipation. To highlight this point we propose to calculate the entropy production rate $\frac{{\rm d}{\Delta S}_{\rm c}}{{\rm d}t}$ associated to the system, which is given by:
\begin{equation}\label{entropiecree}
\frac{{\rm d}{\Delta S}_{\rm c}}{{\rm d}t} = \frac{Q_S}{T_S} + \frac{Q_P}{T_P}
\end{equation}
\noindent Replacing $Q_S$ and $Q_P$ using Eq.~(\ref{eq1}), this expression becomes:
\begin{eqnarray}\label{entropiecree2}
\frac{{\rm d}{\Delta S}_{\rm c}}{{\rm d}t} &=& \frac{1}{T_S T_P} \left[ K \left(T_S - T_P\right)^2  + \frac{T_S + T_P}{2}  \frac{I^2}{G} \right]
\end{eqnarray}
\noindent This form clearly shows that there are only two contributions to the entropy production and thus to the dissipation of energy: the first term is associated to heat conduction while the second term is associated to Joule heating. However the two convective terms appearing in the expressions of $Q_S$ and $Q_P$ does not contribute to entropy production since, once divided by the temperature of each thermal reservoir, they cancel each other: the convective heat flux due to thermoelectric effects may be said reversible. In practice, this reversibility appears because of the fact that a reversal of electrical current leads to a reversal of the convective thermal flux.
It is interesting to note that, in the linear framework of Ioffe's model, the total power dissipation $Q_{\rm Total}$ defined in Ref.~\cite{Lee2013} is actually only related to dissipation since it only depends on Joule heating: 
\begin{equation}
Q_{\rm Total} = Q_P + Q_S = \frac{I^2}{G} = GV^2.
\end{equation}
\noindent Since for the setup used $T_P = T_S$, the heat conduction has no influence on the dissipation.

The description of the reversible Carnot engine gives another argument against the bold identification of heat exchanges with dissipation: In the refrigerator regime, during a whole cycle of this heat engine, some heat is taken from the cold reservoir and then transferred to the hot reservoir \cite{Carnot1824}. However the incoming heat in the hot reservoir cannot be qualified as dissipation since the engine works in a reversible fashion.

\section*{Nonlinearities}
While we have based so far our analysis on the linear model proposed by Ioffe, it is clear that the behavior of the molecular junction can no longer be considered as linear for the values of V considered in Ref.~\cite{Lee2013} as already stressed by Lee and coworkers. However, we believe that the insights highlighted with the preceding linear model could easily be extended to the nonlinear case. In particular, one might suspect that the asymmetry between the two thermal fluxes $Q_P$ and $Q_S$ is always associated to reversible phenomena and hence may not be related to dissipation. For example, in Ref.~\cite{Zotti2014}, Zotti and coworkers consider a single-level model: they demonstrate that the total power contains only even orders in V, and is thus associated to dissipations, while the asymmetry contains only odd orders. As the odd terms regarding applied voltage are characteristic of a reversible behavior (since when a voltage is reversed, the thermal flux is then also reversed), this result supports our hypothesis that asymmetry is not related to dissipative processes. It would be interesting to check if it is always the case independently of the system considered, i.e., if the hypothesis of equipartition of the dissipation between the two ends of a thermoelectric module made by Ioffe is always valid.
\section*{Conclusion}

The asymmetry observed in Ref.~\cite{Lee2013} concerns only heat fluxes and not dissipation.
This feature is associated to the presence of a convective heat flux that is not specific to mesoscopic devices: it is a general property of thermoelectric systems independently of its size.

%\section*{Analysis of the results within the linear approximation}
%
%\begin{equation}
%Q_{\rm Total} = Q_P + Q_S = \frac{I^2}{G} = GV^2
%\end{equation}
%\begin{equation}
%Q_P = \frac{Q_{\rm Total}}{2} \pm S T \sqrt{G Q_{\rm Total}}
%\end{equation}
%

%\begin{acknowledgments}
%Y. A. thanks O. Entin-Wohlman, J.-H. Jiang and Y. Imry for encouraging him to make this comment.
%\end{acknowledgments}

\end{document}